# Clustering based method for finding spikes in insect neurons

Smith Gupta

Indian Institute of Technology, Kanpur

## Abstract

Spikes can be easily detected in most intracellular recordings as sharp peaks. However, in some experimental preparations, because of unipolar morphology or other characteristics of the recorded neurons, the sizes of the spikes recorded from the soma can be much smaller. The experimental settings and the quality of the recording can also affect the observed amplitudes of the spikes. Whole-cell patch-clamp recordings from the somata of projection neurons of the antennal lobe in Drosophila or mosquitoes can show spikes with amplitudes as small as 2 mV. Moreover, the observed spikes often ride on relatively large depolarizations, which makes it difficult for the standard thresholding-based approaches to distinguish them from noise or sharp EPSPs present in the signal. For spike detection in such neuronal recordings, we propose a clustering-based algorithm that separates peaks corresponding to action potentials from those corresponding to noise. Candidate peaks, including many noise peaks, are first selected according to their sharpness, and then a feature vector is extracted for each peak. The 3-dimensional feature vector contains the absolute value of the peak voltage, height of the spike, and the magnitude of the second derivative minima attained during the spike. In most recordings, this 3D space reveals two natural clusters, separating the noise peaks from the true action potentials. Some parameters of the algorithm can be optionally altered by the user to improve detection, which comes handy in the few recordings where the default parameters do not work well. In summary, the algorithm facilitates accurate spike detection to enable the interpretation and analysis of patch-clamp data from neuronal recordings in invertebrates. The algorithm is implemented in an open-source tool, which will be freely available to the community.

## Introduction

In an electrophysiological recording of membrane voltage of a neuron action potentials stand out distinctly as a rapid increase followed by almost equally rapid fall back to the resting potential. Detection of these spikes becomes an essential first step for any kind of analysis to be performed on these recordings. Due to this, a lot of workers have contributed algorithms that keep getting more accurate at detecting spikes. One of the standard methods to detect spikes involves setting a threshold for either the peak amplitude(Laboy-Juárez et al., 2019), the peak sharpness, or both(Haver, 2014). The value of threshold needs to be set a priori for each recording, which becomes a difficult task as in practice it is seen that the proper value of the threshold varies with shape and size of the spikes in the recording. In addition to these, there are algorithms designed to work with extracellular recordings from mammalian neurons(Kim and Kim, 2003; Wang et al., 2016). The shapes of the action potentials recorded with the two types of techniques -- intracellular and extracellular -- differ significantly. This results in poor spike detection when extracellular algorithms are used with default parameters on intracellular recordings, and requires human intervention to correctly set the parameters for each recording. In the case of action potentials recorded from invertebrate neurons with intracellular techniques, like whole-cell patch-clamp, amplitudes can be as low as 10 mV(Wilson et al., 2004; Gouwens and Wilson, 2009). In this technique, a vacuum patch is made on the cell body of the neuron and an electrode is inserted through it, while the other electrode remains in the extracellular fluid(Murthy and Turner, 2013). If the neuron under inspection is a unipolar neuron, which is a common type in the invertebrate nervous system, and spike initiation zone is farther from the cell body then the recorded voltage has a dissipated amplitude(Gouwens and Wilson, 2009; Murthy and Turner, 2013). Figure 1 shows whole-cell patch-clamp recordings from projection neurons in the antennal lobe of Aedes aegypti, and action potentials in these recordings vary from 2-10 mV. Another factor that causes these small amplitudes are the EPSPs generated by the upstream neurons which distort the shape and size of the action potential in these neurons. Also, in such recordings presence of sharp noise and other artefacts can make it difficult to detect the action potentials by thresholding methods alone.

Here we present an algorithm to detect such action potentials that does not require to set any explicit threshold. The first step of the algorithm broadly selects a set of candidate peaks in accordance with their sharpness, and this set contains peaks that correspond to true action potentials and those that corresponds to noise. True action potentials are separated from the noise peak by extracting 3-dimensional feature vectors for each peak and spreading them in 3-D space. The feature vector has absolute peak voltage, peak height, and the absolute value of the second derivative minima as its elements. These points are then scattered and separated into two clusters by means of hierarchical clustering. These features separate out the noise peak on the lower corner of the space, while the true action potentials cluster on the opposite corner (Fig. 6).

The source code is made freely available at [link](). The features used here are heuristically chosen such that they correlate with the distinct shape of the action potentials observed. These can be altered as per the expectation of spike shapes which would depend on the neuron and the recording technique employed on it. This results in a general solution for spike detection in intracellular recordings that gives high accuracy with minimal requirement of human intervention.

## Methods

### Software

Matlab R2019b was used for all data analysis. Low-pass filtering was performed with a third-order Butterworth filter using the functions 'butter' and 'filtfilt' in Matlab.

### Recordings

The whole-cell patch clamp recordings that are used here to validate the algorithm come from different sources. The recordings from *Aedes aegypti* olfactory projection neurons is collected by Pranjul Singh and Shefali Goyal as part of an ongoing work. The recordings from *Drosophila melanogaster* are acquired from previously published studies that describe the roles of olfactory projection neurons(Shimizu and Stopfer, 2017) and Mushroom Body Output Neurons(Hige et al., 2015).

# Results

## Recording Examples

In figures 1a-f we show whole-cell patch-clamp recordings from olfactory projection neurons of *Aedes aegypti*. The recordings are chosen to showcase variations in spike sizes and firing rates that are generally encountered in this setup. We are going to look at the limitations of traditional algorithms when they are applied to these recordings, followed by the description and results of our proposed method.

## Detecting spikes with a second derivative-based threshold

One of the widely used algorithms for spike detection is based on using a threshold for second derivative values of the membrane potential, and marking those peaks as action potentials that cross this threshold value during the rise time or the fall time. The threshold is calculated using the standard deviation of the second derivative values, during the background interval, multiplied by a constant factor.

$$Threshold = k * \sigma_{V''_{bi}}$$

A generally used value for k is 2-3, as most of the action potentials are expected to lie within three standard deviations. But using a fixed value of k for a set of recordings is prone to errors, as a particular value of k can give high false positive rate for one recording while giving low true positive rate in the other (see Fig. 2). So more than often this value has to be set manually depending on the individual recording. Moreover, there can be recordings where there is no sharp divide between second derivatives values in the action potential regions compared to those in the noise regions. For these cases, even after setting individual thresholds manually for each recording, the true positive rate and false positive rate do not hit 1 and 0 together.

## Detecting spikes with an amplitude-based threshold

Another algorithm to detect spikes relies on setting a threshold on the amplitude of each peak measured from a baseline. The baseline is defined as the average membrane potential during the pre-stimulus period.

$$V_{baseline} = mean(V_{bi})$$

Now for each peak, we check if the height from the baseline to the peak is greater than the set threshold to mark it as a spike.

$$V_{max} - V_{baseline} \geq threshold$$

Figure 3 shows such an algorithm applied to recordings of fig 1d and 1e. From the ROC curve it can be seen that for the amplitude based algorithm also no particular threshold value works for all the recordings, and even with manually set thresholds for each recording we get poor true positives with high false negatives.

## Three features for spike characterization

As the first step for peak identification, we look at all the zero-crossing points of the first derivative of the signal, and on this set we use a loose second derivative threshold to narrow down to a set of candidate spikes. Using a value of 0.5 for k in equation 1 ensures that we get all true action potentials as our candidate spikes, which will be accompanied by a lot of noise peaks as well. The signal should be smoothed out before identification of candidate peaks, this will ensure minimal noise peaks appearing in our set. Figure 4a shows a 100 ms snippet of the raw signal from recordings of Fig 1a and 1d. On applying the second derivative thresholding on the raw signal itself, we get a lot of noise peaks detected along with true action potentials as shown in Fig 4b. A low-pass filter smoothens these noise peaks, and a cut-off frequency can be chosen to do this without distorting the shape of action potentials. For a sampling rate of 20 kHz, a frequency of 500Hz for the low pass filter works well. Fig 4c shows the signal after the application of this low-pass filter, and Fig 4d shows the candidate spikes on the filtered signal. After filtering, the true action potentials remain as is in the set of candidate spikes, with a huge reduction in the number of noise peaks.

Once we have a set of candidate peaks that contain true action potentials along with noise peaks, the next goal becomes to separate these two sets. For this task, we extract a 3-dimensional feature vector for each candidate peak. The feature vector should capture information about the sharpness of the peak, so as a first element we take the absolute value of the second derivative minima that is reached during the rising time or the falling time of the action potential. For the second element, the height of the peak is calculated as the minimum of the two heights of the peak from the troughs of either side. In cases where the action potentials are riding on EPSPs, the heights are less but the voltage values at the peak are high. So peak voltage becomes an important feature for identification of such action potentials and is taken as the third element of the feature vector (Fig. 5).

Features

$V_{max}$ $\longrightarrow$ maxima of the membrane potential

$\Delta V$ $\longrightarrow$ max($\Delta V_1$, $\Delta V_2$)

$|V''_{min}|$ $\longrightarrow$ absolute value of second derivative minima

## A clustering-based algorithm for detecting spikes using the three features

We used the normalized value of each vector and on scattering these values in 3d space two natural clusters appear (Fig. 6). On marking the points back on the signal after cluster labelling, we see that all the true spikes indeed cluster together with the defined features.

Fig. 7 shows all the recordings from fig. 1 with action potentials marked on them. The true positive rate in each case is 1 with 0 false-positive rate, and the algorithm works as well as a human expert marking spikes with visual inspection.

We also verified this algorithm on datasets containing recordings from Drosophila's antennal lobe projection neurons(Shimizu and Stopfer, 2017) and mushroom body neurons(Hige et al., 2015). The dataset from the antennal lobe contained six neurons from each of the four glomeruli – DL2v, VC3, VC4 & VM5v, and the odors used were benzaldehyde, 2-octanone, ethyl acetate, pentylacetate, and ethyl butyrate. The total number of true action potentials in all the recordings from this dataset were 75,422, out of which our algorithm detected 75,412 leaving out only 10 spikes with 2 false positive spikes (fig. 8). From the mushroom body we used data from two Kenyon cells, and seven cells from each of the two MBON types -- MBON-$\alpha$1 and MBON-$\gamma$1pedc. The odors used were 3-octanol, 4-methyl cyclohexanol, and 2-heptanone. Membrane potential traces from the two Kenyon cells contained 37 spikes in all, and our algorithm successfully detected all of these. Our algorithm fared worst for the mushroom body output neurons dataset where there were 26,605 spikes in all, out of which we accurately detected 26,426 spikes, leaving out 179 spikes with 35 false positives (fig. 8). This was due to high noise levels in these recordings, and it is difficult to correctly mark many of these peaks even with visual inspection.

## Discussion

When recordings are done on certain classes of neurons the action potentials can have widely varying shapes and sizes. This poses a problem for methods that involve setting up a fixed threshold on any recording parameter, and the use of these methods require significant human intervention in fixing the appropriate values for the thresholds used. Here, we have discussed a method that circumvents this issue by employing relative feature differences between the shapes and sizes of noise peaks and the true action potentials. The features chosen here are from the perspectives of patch-clamp recordings, and we extract information relevant for detecting action potentials that are usually seen in these recordings. However, depending on the recording technique a user can change these features to accommodate any other particular property of the action potential that can differentiate it from the noise peaks. The source code for the method is available online and has been made user-friendly to help setting the right parameters for various recordings. We hope this method lessens the human intervention required in the task of spike detection and reduces efforts for workers doing electrophysiological recordings.

Figure 1 | Whole cell recordings from *Aedes aegypti*
Whole-cell patch clamp recordings are shown from six antennal lobe *projection neurons* of *Aedes aegypti*. The recording duration for each is 10s, and odor stimulus period is from 2-3 seconds (shown with the bar). The amplitude of spikes in these recordings are as low as 2 mV.

Figure 2 | Performance of second derivative-based algorithms
ROC curves for the results produced by second derivative-based spike detection algorithm on recordings taken from fig. 1 (cell 1 and cell 5), along with a one second snippet of the results for three values of k: 1.5, 2, and 2.5. (a) For the recording from cell 1, all the spikes get detected for the value of k less than 1.7 but with a false positive rate higher than 0.4, and all the false positives are removed only for a value of k that is at least 3.6 but with a maximum true positive rate of 0.965. (b) For the recording from cell 5, a value of k greater than 1.2 starts eliminating true spikes, and false positive remains till the value of k is below 2.

Figure 3 | Performance of amplitude-based algorithms
ROC curves for the results produced by amplitude based spike detection algorithm on recordings taken from fig. 1 (cell 4 and cell 5), along with a one second snippet of the results for three values of k: 1, 2, and 3. (a) For the recording from cell 4, threshold needed to encompass all the spikes is lower than or equal to 0.8, and a threshold value of 2 leaves out many of the true spikes giving a high false negative rate. (b) For the recording from cell 5, threshold needed to eliminate all the false positive spikes is 11.1, and a threshold value of 2 leaves out a lot of noise peaks marked falsely as spikes giving a high false positive rate.

Figure 4 | Low pass filtering
100 ms snippet of recordings from Cell 1 and Cell 4. (a) Raw recording trace. (b) Candidate spikes detected using the raw trace after applying a second derivative threshold of 0.5**sd*. (c) Filtered recording trace with low pass frequency of 500 Hz. (d) Candidate spikes detected using the filtered recording trace after applying a second derivative threshold of 0.5**sd*.

Figure 5 | Features for clustering
Illustration of feature vector components used for separating noise peaks from action potentials. The three features used are the maxima of membrane potential (Vmax), max of the two spike heights from either side of the peak ($\Delta V_1$ and $\Delta V_2$), and the absolute value of second derivative minima in the rising or falling time of a peak.

Figure 6 | Cluster separation of noise peaks and action potentials
Candidate spikes from recordings of Cell 1 and Cell 4, marked and separated into two clusters. (a) The three features are plotted in a 3D space, each point represents a peak in the signal. Cluster separation is done with a hierarchal algorithm, action potentials cluster on the higher end of the cube (marked with blue), while the noise peaks cluster in the lower corner (marked with red).

Figure 7 | Final result
Recordings from the six cells are shown here with the final detected spikes after cluster separation.

Figure 8 | TPR and PPV on Drosophila datasets
Performance of algorithm shown with the metrics of True Positive Rate (TPR) and Positive Prediction Value (PPV) on neurons from Drosophila olfactory system. The recordings are taken from antennal lobe projection neurons (VC3, VC4, DL2v, VM5v), mushroom body output neurons (MBONa1, MBONg1), and Kenyon cells (KC).

Figure 1

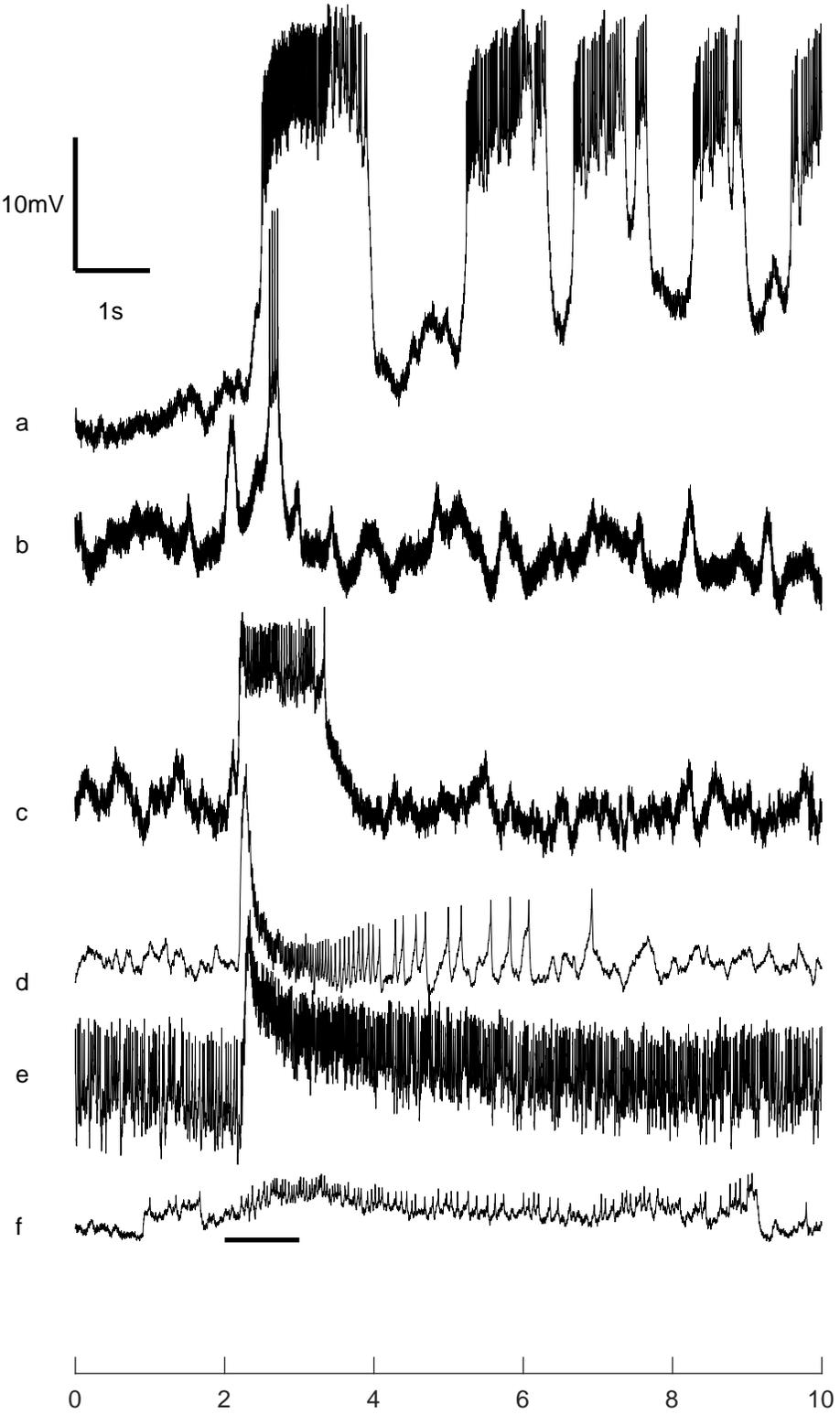

Figure 2

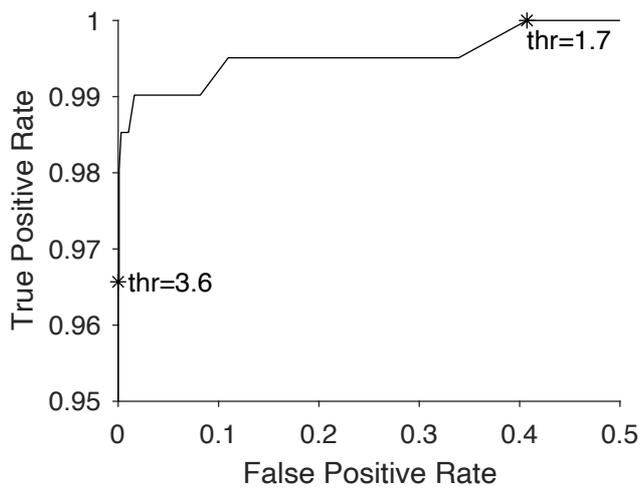
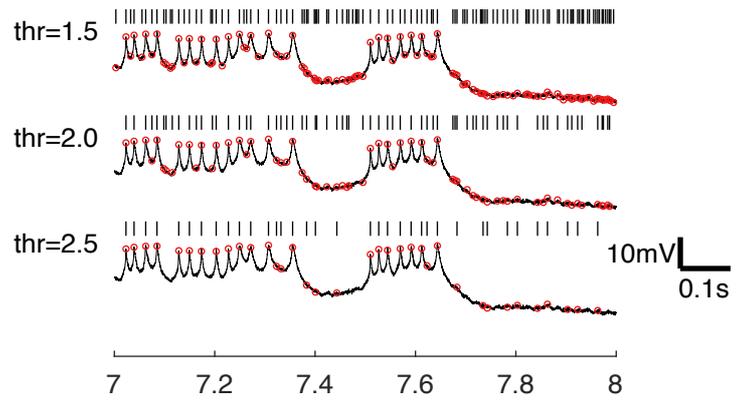
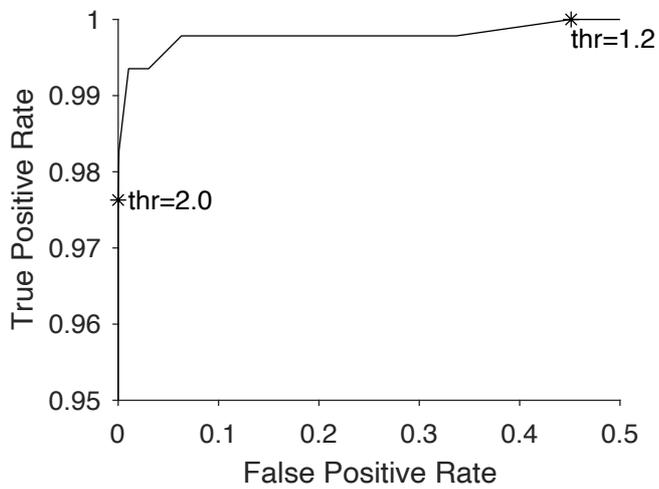
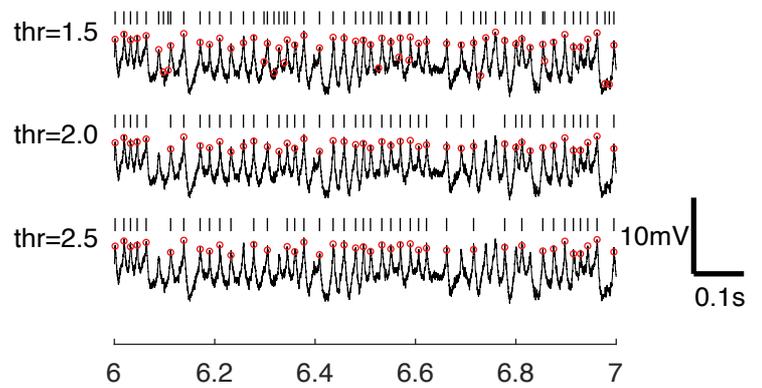

Figure 3

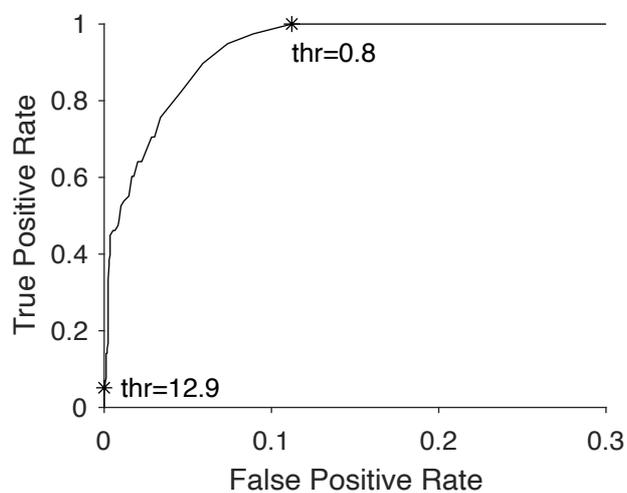
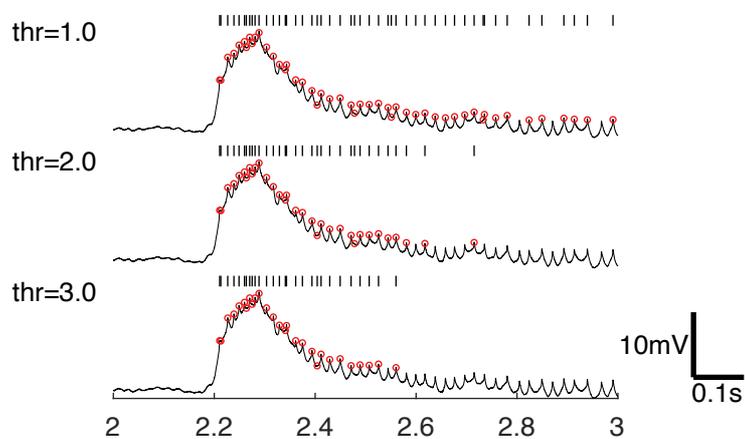
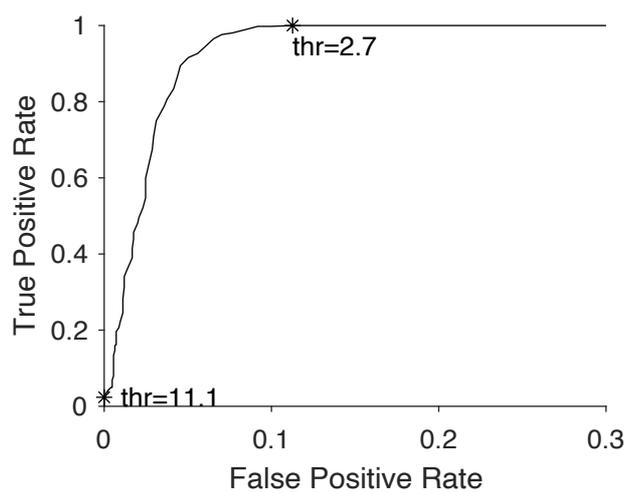
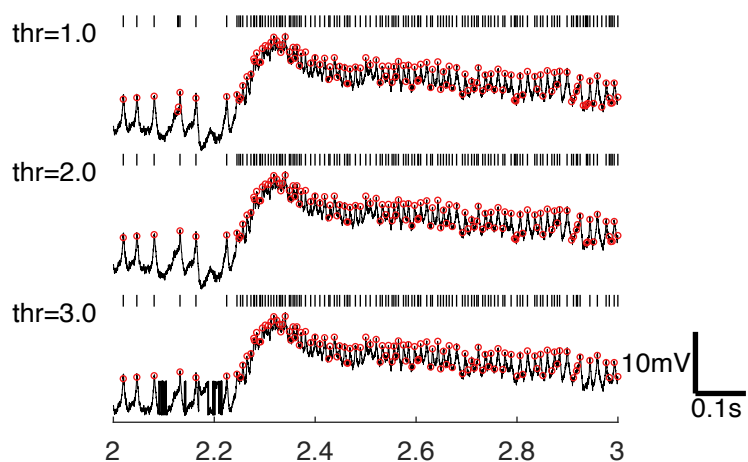

Figure 4

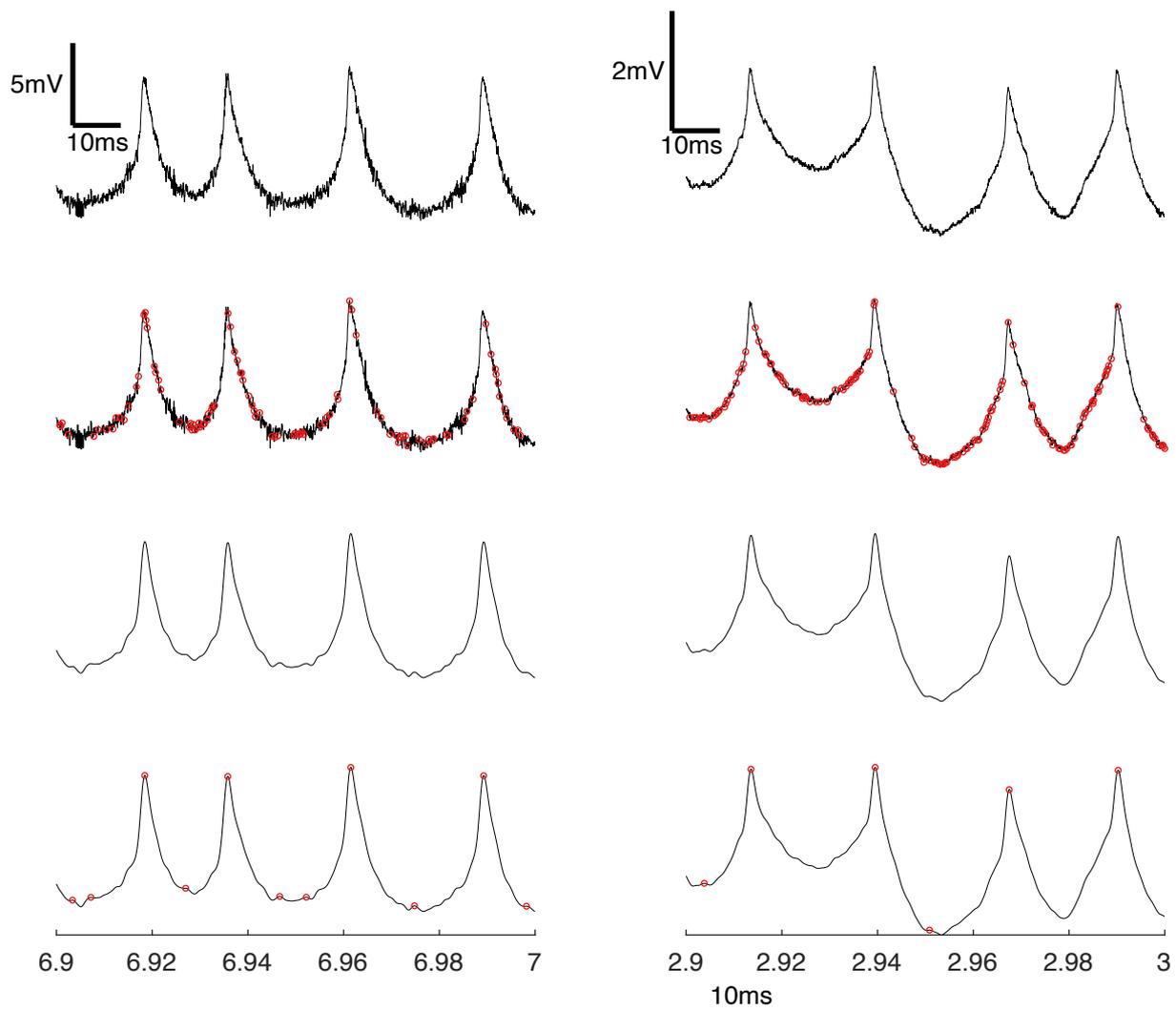

Figure 5

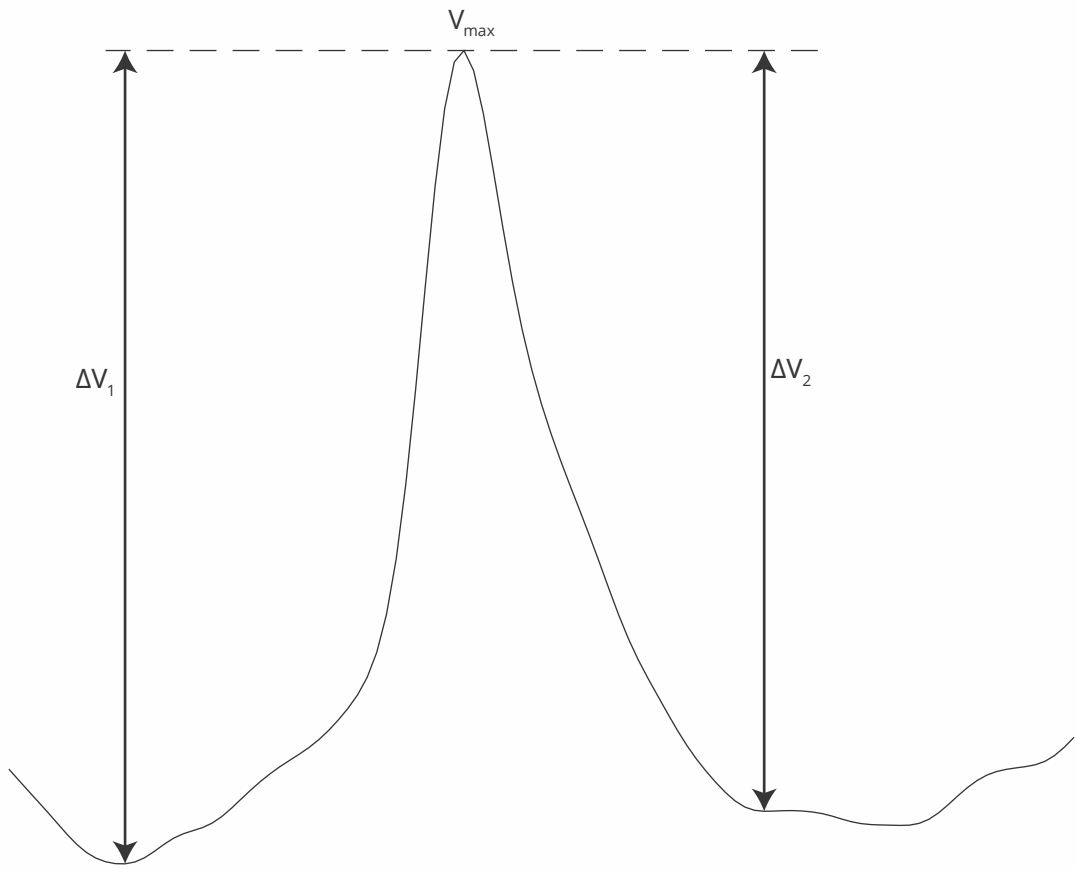

Figure 6

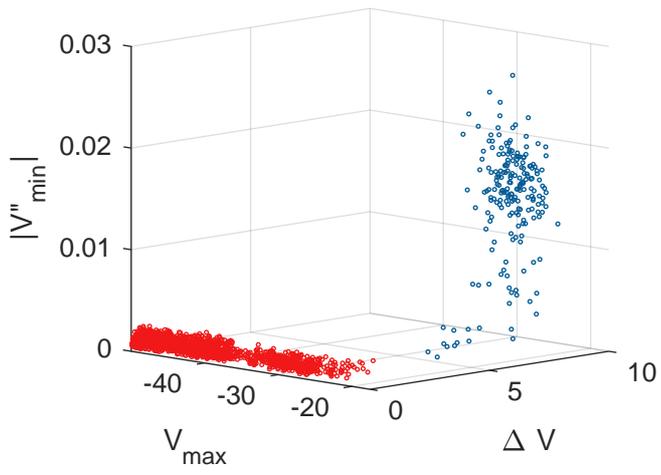
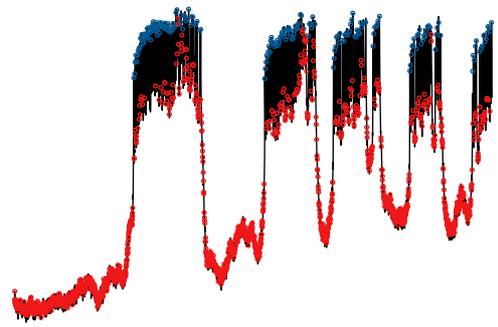

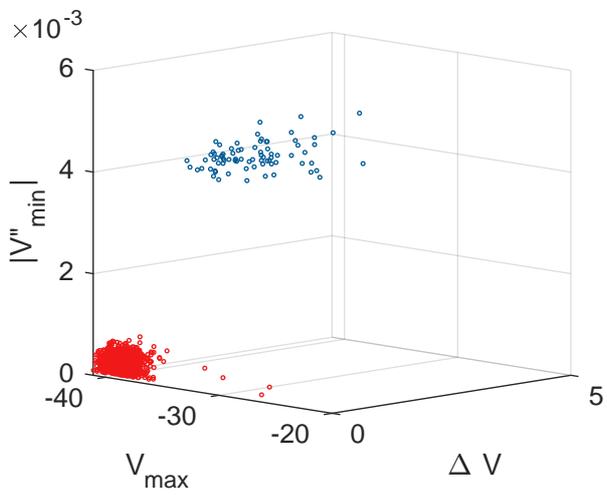
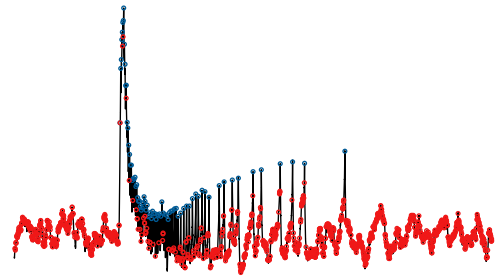

Figure 7

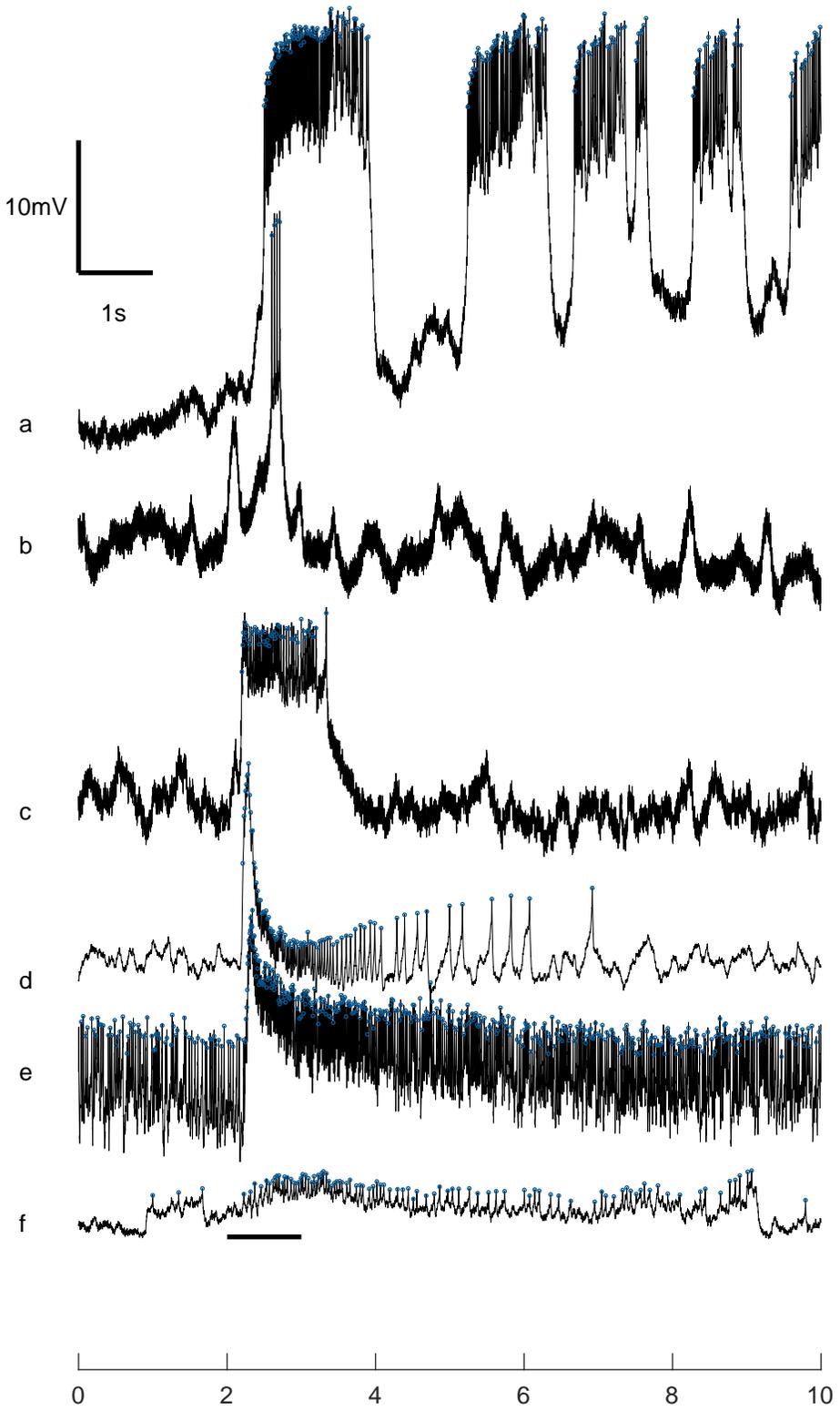

Figure 8

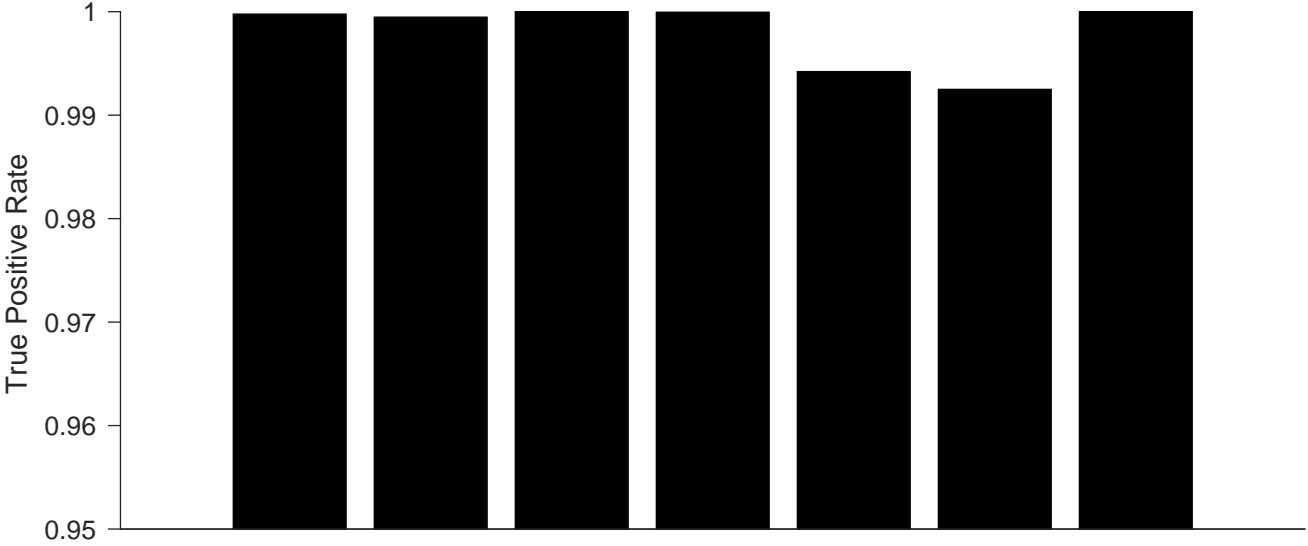
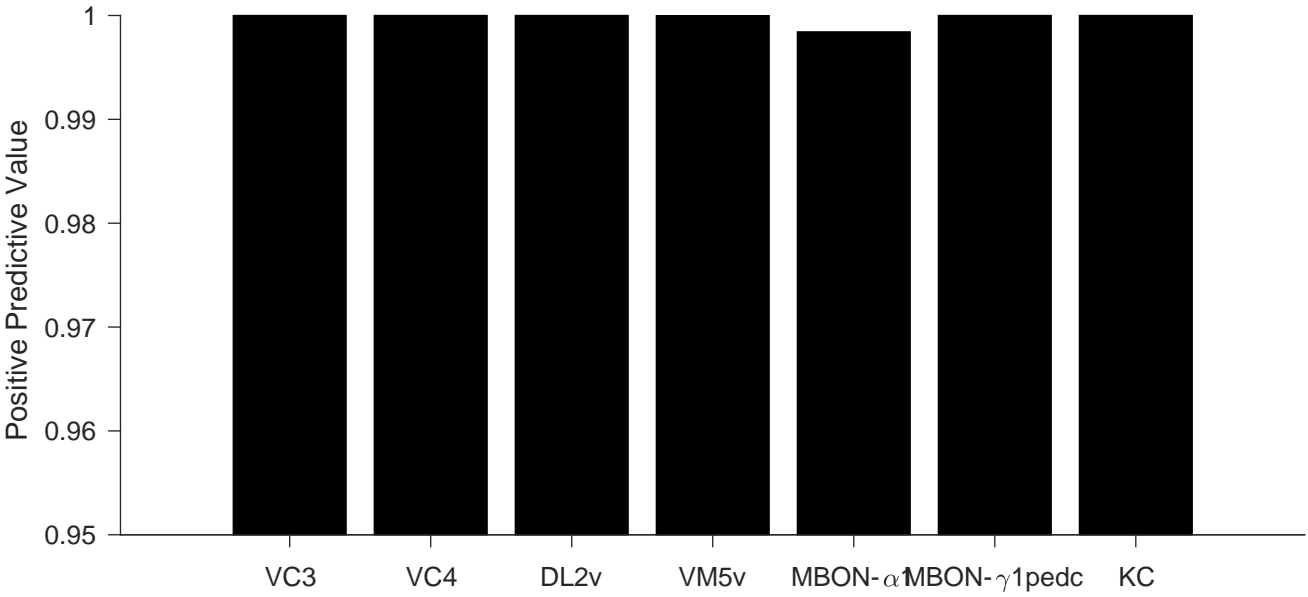